# Color Segmentation on FPGA Using Minimum Distance Classifier for Automatic Road Sign Detection

Jingbo Zhao, Benny Thörnberg, Yan Shi, Ashkan Hashemi
Department of Information Technology and Media
Mid-Sweden University
Sundsvall, Sweden
jizh1000@student.miun.se, benny.thornberg@miun.se, yash1000@student.miun.se, hashemi.ashkan@gmail.com

*Abstract*—Classification is an important step in machine vision systems; it reveals the true identity of an object using features extracted in pre-processing steps. Practical usage requires the operation to be fast, energy efficient and easy to implement. In this paper, we present a design of the Minimum Distance Classifier based on an FPGA platform. It is optimized by the pipelined structure to strike a balance between device utilization and computational speed. In addition, the dimension of the feature space is modeled as a generic parameter, making it possible for the design to re-generate hardware to cope with feature space with arbitrary dimensions. Its primary application is demonstrated in color segmentation on FPGA in the form of efficient classification using color as a feature. This result is further extended by introducing a multi-class component labeling module to label the segmented color components and measure their geometric properties. The combination of these two modules can effectively detect road signs as the region of interests.

## I. INTRODUCTION

The structure of machine vision systems [20] typically consists of image acquisition, pre-processing, segmentation, post-processing, labeling, feature extraction and classification. Practical implementation of the segmentation stage partitions an image into the foreground objects and backgrounds by thresholding. While this method may work for simple tasks such as the segmentation of coins from a table surface, in a more complex scenario, when different colors are sharing similar grayscale, it is likely to fail. Therefore, devising a solution for multi-color segmentation that is specifically suitable for the FPGA structure is eagerly needed.

The color segmentation stage can also be viewed as a feature classification process, if the color components are considered as features and a classifier is used to classify the feature vector of colors. The problem arises; when a classical classification algorithm such as Naïve Bayesian Classifier is to be implemented on FPGA, for the use of the Gaussian function is too cumbersome to be realized on hardware. A simple and efficient method is preferable in real-time applications.

The Minimum Distance Classifier [8][9] is a method that classifies feature vectors based on selecting the shortest distance of an input vector with respect to all class centers and assigning the vector to the class center that reveals the shortest distance. It is a simple and decent way to classification problems. The distance between two vectors is often evaluated based on the different forms of the norms, such as the Manhattan distance and the Euclidean distance. To obtain the class centers, normally it is possible to directly perform a mean value calculation on data sets with known classes or alternatively by applying training algorithms such as K-means or Mean Shift.

## II. RELATED WORK

In retrospect, there have been many works conducted on segmentation using color as a feature. Examples are Normalized Cuts [2] that performs the segmentation in a geometrical approach; K-means as a clustering algorithm for finding clusters with similar features; Self-organizing Map as an extended version based on K-means and has evolved into a form of neural networks [8]; and Mean shift, which is also a clustering algorithm based on finding modes in data vectors. But all of these methods are non-deterministic and iterative or contain non-linear functions. The implementations of these algorithms, although may exist, take a lot of hardware resources. Thus, the Minimum Distance Classifier with no non-deterministic and iterative operations in a linear form is a more beneficial option for color segmentation on FPGA since they potentially take fewer hardware resources.

On the other hand, systems that use the Minimum Distance Classifier have already appeared. For example, in [22], the proposed method is used for automated kiwi counting based on its color profile. In [21], the method uses the Minimum Distance Classifier for the shape classification of the traffic signs. However, these are software methods that have not been implemented on hardware.

Specifically, for road sign detection systems on FPGA, it seems that the Minimum Distance Classifier has not been used for color segmentation before. Alternatively, a system based on using the feature of the Histogram of Orientated Gradients (HOG) has been proposed and implemented for pedestrian detection [14] and it has been extended for the detection of the stop sign [15]. But this method relies on the calculation of the integral map (IMAP) that contains the HOG features. One bin in the histogram requires one frame buffer to hold it. Thus, it consumes a large amount of memory for buffers. Moreover, it needs to perform block processing on the frame buffers at various scales to re-generalize the HOG from IMAP, which is also a point that makes it inefficient.

To the authors' knowledge, there has not been any result on using a Minimum Distance Classifier module on FPGA to segment the color images and using a hardware multi-class labeler to measure the properties of the segmented components for road signs detection.

## III. METHODOLOGY

The structure of a machine vision system that has been introduced in the beginning is also valid for the road sign detection system proposed in this paper. It can be generalized as shown in Figure 1. A detailed explanation is given in this chapter and a particular emphasis is placed on the kernel of the Minimum Distance Classifier.

### A. Image Acquisition

Industrial cameras may directly capture images in YCbCr format. By removing the Y component and keeping Cb and Cr components for color segmentation, it is possible to decouple the interference of lighting conditions that may affect its performance. This is more advantageous than performing such an operation in the RGB format. Since in this color space, the result can be heavily affected by illumination. Alternative color spaces that can be used are CIE-LUV and CIE-LAB, which are device-independent and are linear spaces for color segmentation, but it consumes much more hardware resources and computational power to perform non-linear transformation from the RGB color space to CIE-LUV or CIE-LAB. Thus, we prefer the YCbCr model since they are directly available. In the proposed system, only Cb and Cr channels are used.

### B. 3 by 3 Gaussian Filter

The captured image may have Gaussian noise; this is very prohibitive for the Minimum Distance Classifier since the pixel with small abnormal color variations could fall out of the decision boundary of the class it should belong to. In this way, it may produce many meaningless small components. In order to label these components, the Multi-class Component Labeler may consume much memory. By introducing a 3 by 3 Gaussian Filter to smooth the image in Cb and Cr channels, these small components are reduced in large numbers.

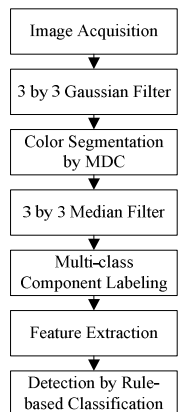

Figure 1. Structure of the Road Sign Detection System

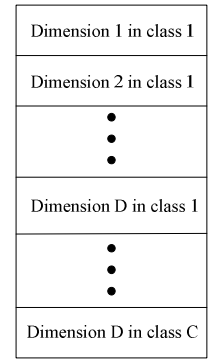

Figure 2. Structure of the Register File

### C. Color Segmentation By Minimum Distance Classifier

*1) The Algorithm for Minimum Distance Classifier:* The algorithm for Minimum Distance Classifier can be generalized as:

For each data point $x_i$, compute the distance to each class center and assign the data point to the nearest class center with the shortest distance.

$$d_i = \min_j d(x_i, u_j) \qquad (1)$$

The term $d(x_i, u_j)$ is the L-norm equation i.e. the distance metric, which can be expanded as:

$$d(x_i, u_j) = (\sum_{i=1}^{n} |x_i - u_j|^p)^{1/p} \qquad (2)$$

On FPGA, since it is needed to economize the hardware resources, the multipliers should be avoided in this case.

By letting *p=1*, the Manhattan distance metric is obtained:

$$d(x_i, u_j) = \sum_{i=1}^{n} |x_i - u_j| \qquad (3)$$

Equation (3) is the distance metric that is used in the implementation of the Minimum Distance Classifier which is free of multiplication operations.

*2) The structure for Minimum Distance Classifier:* The operation of the minimum distance classifier has two stages: first programmed with class centers by another module such as a test bench into the register file and then it performs classification operations with given inputs. The class centers are stored in a one-dimensional array, with each cell containing one dimension of one class center vector. This allows easier calculation of the addressing index in practice. We define the number of classes as C and the number of dimensions as D. With this definition, the register file is arranged as shown in Figure 2.

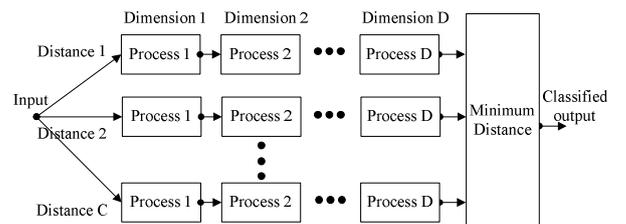

Figure 3. Structure of the Classification Stage

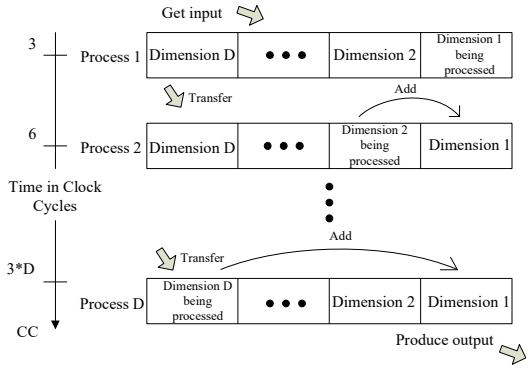

Figure 4. Buffer of the Pipeline

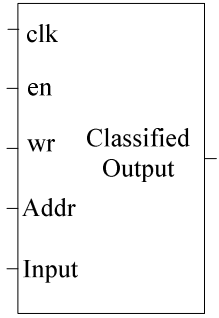

Figure 6. Interface of the Minimum Distance Classifier

The indexing of the register file is provided by an address port, which is explained in the interface part.

In the classification stage, the operation can be further generalized into two levels, one for the calculation of the distance and the other for the process of selecting the shortest distance.

As shown in Figure 3, the distance calculation is performed in the number of C parallel pipelines. Each input is calculated with respect to every class center, so that the results can reach the minimum distance module at the same time.

To make each distance calculation in a pipelined manner, a buffer that corresponds to the input vector is used. The number of cells is the same as the number of the dimension of the input vector.

An explanation of the buffer is that in the first process (Input stage) of the pipeline, the input vector is read in and the first dimension is processed. Then the vector is transferred to the second process (Intermediate stage) and the second dimension is processed. It should be emphasized that the result is added to the first dimension. In the last process (Output stage), the dimension is processed and the result is also added to the first dimension so that when selecting the shortest distance, only the first dimension of the result is used. The result of the process is transferred to the minimum distance module. Each process takes 3 clock cycles. By increasing the dimensions, the number of intermediate stages is increased automatically to generate new hardware.

The module that selects the shortest distance can be further explained by Figure 5, where a comparison of values by pair is performed. For cases there are no candidates available to make pairs, it is delayed for a number of clock cycles as is characterized by Distance C.

*3) The interface for Minimum Distance Classifier*

The interface of the minimum distance classifier resembles an interface of a memory module, in such a way that the classifier can be programmed with class centers by test bench or another module and can perform classification operations with given inputs. The idea is fulfilled with the interface in Figure 6.

The signal *clk* is clock pin and *en* is the enable pin. When the write signal *wr* is high, the address port *addr* provides the index to store class centers in the register file. When *wr* goes low, the label port labels can produce output with given input at the *input* port.

The resolution of a single dimension in *input* port is R. To account for the overflow that may occur during calculation, two extra bits should be added. The overall *width for all dimensions* of input is $(R + 2) * D - 1$ bits. The width of *addr* port is ceiling of $\log_2(D * C) - 1$ bits and the width of classified output is the ceiling of $\log_2 C - 1$ bits.

### D. 3 by 3 Median Filter

The segmented image contains corrupted objects such as numbers on the road signs resulting in too many meaningless components. These still put burden on the memory of the

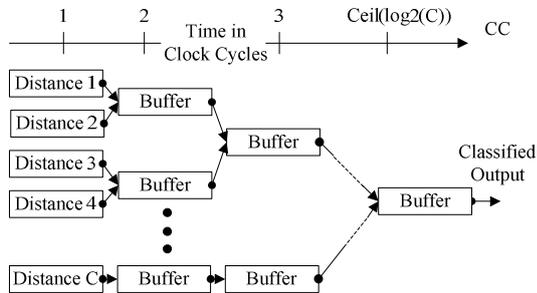

Figure 5. Structure of the Minimum Distance Selection Module

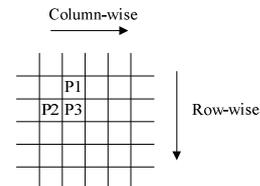

Figure 7. 4-connectivity Neighborhood for Labeling

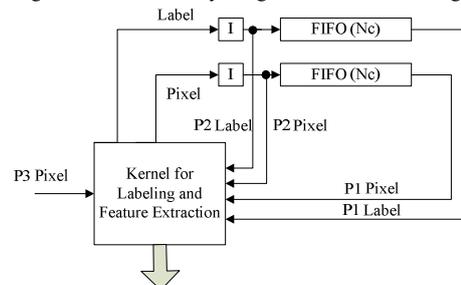

Figure 8. Labeling Kernel and Line Buffer

Multi-class Component Labeler and it is difficult to perform a recognition algorithm on the corrupted number. The solution is to use a 3 by 3 Median Filter that further reduces the small components based on the results achieved by the 3 by 3 Gaussian Filter. This repairs the useful components while retaining the original segmented image.

*E. Multi-class Component Labeling and Feature Extraction*

Conventional labeling methods cannot work on segmented multi-class images. Previously, at Mid-Sweden University, a kernel for component labeling and feature extraction has been developed [7][11]. By adding an additional line buffer for buffering the input pixels beside the labels and taking account of the pixel value when assigning the labels, the original kernel can be used for Multi-class component Labeling.

Figure 7 and Figure 8 present the pixel stream running direction and the structure of the buffer of the kernel for Multi-class Labeling.

*F. Detection by Rule-based Classification*

With the processing steps that have been carried out, it is easy to know the area of the labeled color components and the color profile of the components. Thus, in this implementation, a color component that is in yellow, with a ratio of width versus height larger than 0.7 and smaller than 3, and with an area larger than 200 pixels is considered a road sign. The reason to include the ratio of width versus height is to avoid the inner part of a zero being considered a road sign.

## IV. RESULTS

The initial experiment of the design was conducted on the software simulation of the VHDL module.

For the test data, a total of 78 images were taken for road signs under different illumination conditions and down-scaled to 1000*630 by the bilinear interpolation.

*A. Obtaining class parameters*

As has been discussed in the introduction part, various methods can be used for obtaining the class means or centers for the Minimum Distance Classifier. But since manual selection of the color in an image does not reflect the statistical distribution of the colors and using K-means requires guessing the number of clusters in an image, we chose to run the Mean Shift algorithm in Matlab over an image to obtain the number of clusters in it and their centers with an empirical bandwidth of 0.4.

Table I Means of Classes

| Label | Background | Yellow | Red | Red |
|---|---|---|---|---|
| Cb | 127 | 88 | 116 | 109 |
| Cr | 128 | 151 | 157 | 180 |

There are two red values in the table, which are used for compensation of the variations of the red color under different illumination conditions.

*B. Reduction of Small Components*

Figure 9 and Figure 10 show the relationship of the number of components with respect to the absence of both Gaussian and Median Filter and the presence of both Gaussian and Median Filters. Results show that by applying both the Gaussian Filter as pre-processing and the Median Filter as post-processing respectively, the number of generated components can be minimized by approximately 95%.

*C. Device Utilization and Power Calculation of the Minimum Distance Classifier Module*

The evaluation of the VHDL module is in terms of the device utilization, power consumption and maximum running frequency based on the Xillinx Spartan 3E xc3s500e-4pq208 platform running at 50 MHz. The VHDL module was synthesized and post-route simulation was performed. The parameters are calculated using the Xpower software in the Xilinx ISE tool set.

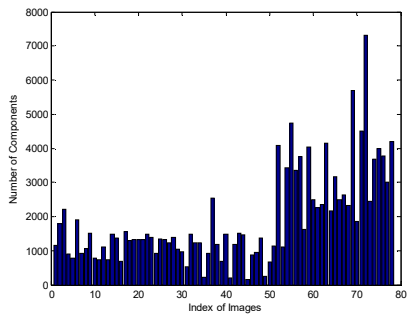

Figure 9. The Number of Generated Components without Gaussian Filter and Median Filter

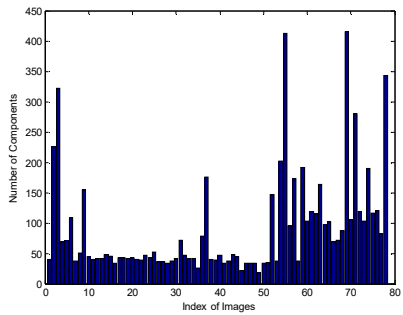

Figure 10. The Number of Generated Components with Both Gaussian Filter and Median Filter

Table II Device Utilization and Power Calculation

| Slices | Slice Flip-flop | 4 Input LUT |
|---|---|---|
| 660 (14%) | 716 (7%) | 895 (9%) |

| Dynamic power (mW) | Clock | Signals | Logic | IOs |
|---|---|---|---|---|
| 0.24 | 0.07 | 0.06 | 0.1 | 0.01 |

| Quiescent Power (mW) | 81.38 |
|---|---|
| Dynamic Power (mW) | 0.24 |
| Total Power (mW) | 81.62 |
| Max Frequency (MHz) | 170 |
| Latency (Clock Cycles) | $3*D+ \text{Ceil}(\log_2 C)$ |

Table II presents the results of the evaluation. For the latency of the design, every dimension in the distance calculation takes up 3 clock cycles, thus the total latency for all of the dimensions takes up 3*D clock cycles. The minimum distance selection part is performed by pair comparison and the latency is Ceil($\log_2 C$), where D is the number of dimensions and C is the number of classes. These two terms together give the total latency that is 3*D+ Ceil($\log_2 C$). Since the maximum running frequency of the design is 170MHz, considering that the resolution of the image is 1000 by 630, the frame rate is approximately 271FPS without taking into account the frame and row synchronization.

*D. Detection Rate*

Table III Detection Rate

| Results | Successful Rate |
|---|---|
| True Positive | 90% |
| False Positive | 2% |
| False Negative | 8% |

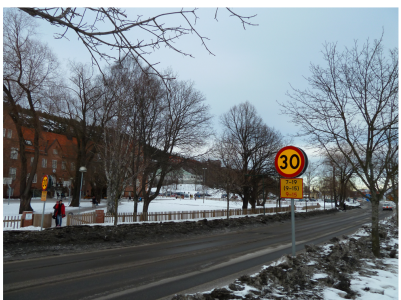
(a) Input Image

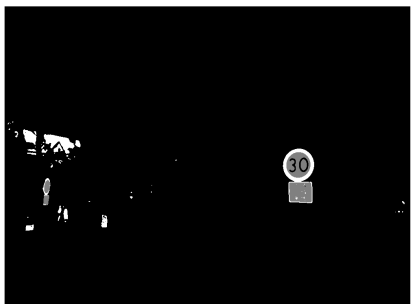
(b) After Multi-Color Segmentation

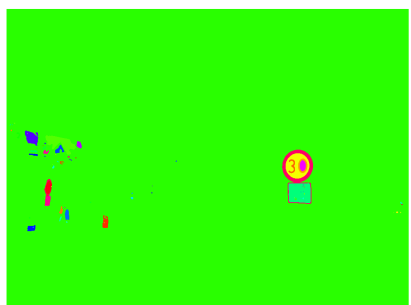
(c) After Multi-class Labeling

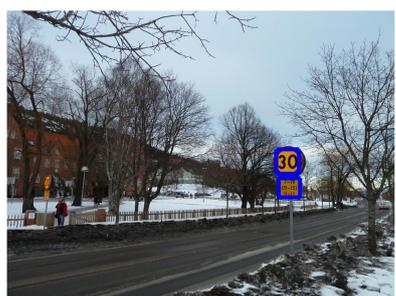
(d) Detected Road Sign
Figure 11. Demonstration of a Detection Example

Table III provides a detection rate for the algorithm out of all 78 images. True positive means the road signs have been successfully detected; false positive means an object, which is not a road sign, has been detected; and false negative means that an object, which is a road sign, has not been detected.

The false positive cases were caused by the narrow angle between the camera and the road sign that has made the road sign deformed too much, causing it shares the same ratio of width versus height with the zeros on the road signs, which are not considered by the algorithm. However, such cases are rare in real situations, since the cameras on the cars are only responsible for detecting road signs straight ahead. These cases may occur when the car reverses and makes U-turns.

False negatives occurred when yellow buildings have been detected as road signs. These cases also do not depreciate the detection algorithm, since in the stage of road sign recognition of the numbers, it can be decided if there are numbers contained in the Region of Interests by calculating the area of the symbols.

*E. Detection Results*

Figure 11 (a) - (d) demonstrates an example of how the road sign is detected. When the input image has been processed by color segmentation, the image has been transformed into a grayscale image with each gray level corresponding to a unique color. After this, the components are labeled by the multi-class labeling algorithm; each of the components is assigned a unique grayscale. Figure 11 is rendered in colors for demonstration purposes. With the labeled image and the rule-based classifier, the road sign is detected and located in the final image.

## V. DISCUSSION

There have been a few deficiencies in the current design of the road sign detection system.

First of all, current testing images were taken in static conditions using a high-resolution digital camera. On real cars, the image could be blurred in low quality due to high speed, thus it cannot be as sharp as it is shown now. A number of 78 images are still not enough. It would be better if it is possible to choose a few industrial cameras that could be potential candidates for the system and put them on a car, running and taking a huge number of images in different road conditions and lighting conditions for testing purposes.

This method is still content-dependent. We have also tried to run it on some of the open datasets available on the internet. It turns out that for those images, it is very difficult to produce a reasonable result using Cb and Cr components. For them, it is much better to use the H component in HSV color space.

As can be seen, by using the current rule-based classifier, the successful detection rate is not high enough. Although explanations have been given for the failing cases, the main problem responsible for this is that the rule-based classifier is too simple. This part can be improved by finding both the outer red ring and inner yellow part and considering the findings of both components as a successful detection of the road sign. In this way, the successful detection rate would increase.

Apart from all these, it would be very promising to incorporate a recognition module, such as something for robust template matching, to recognize the numbers on the road sign so that the system consists of both detection and recognition parts.

Finally, after a proper camera is chosen and the corresponding circuit has been designed, the most interesting thing is to implement the design on a car and observe its real-time performance on road.

## VI. CONCLUSIONS

We have presented a design of the minimum distance classifier on the FPGA platform and demonstrated its application in road sign detection. Results show that the kernel of the Minimum Distance Classifier meets our design requirement. However, at the same time, it would take much more efforts to implement the whole system to let it work in real time.